\documentstyle[12pt,epsf,epsfig]{article}
\def\gsim{\mathrel{\rlap {\raise.5ex\hbox{$ > $}}
{\lower.5ex\hbox{$\sim$}}}}
\def\lsim{\mathrel{\rlap {\raise.5ex\hbox{$ < $}}
{\lower.5ex\hbox{$\sim$}}}}

\newcommand{\pr}{\paragraph{}}
\newcommand{\be}{\begin{equation}}
\newcommand{\ee}{\end{equation}}
\newcommand{\bea}{\begin{eqnarray}}
\newcommand{\nn}{\nonumber}
\newcommand{\eea}{\end{eqnarray}}
\newcommand{\nd}[1]{/\hspace{-0.6em} #1}

\def\gappeq{\mathrel{\rlap {\raise.5ex\hbox{$>$}}
{\lower.5ex\hbox{$\sim$}}}}
 
\def\lappeq{\mathrel{\rlap{\raise.5ex\hbox{$<$}}
{\lower.5ex\hbox{$\sim$}}}}
 
\begin{document}
 
\begin{titlepage}
\begin{flushright}
ACT--6/98 \\
CTP-TAMU--21/98 \\
OUTP--98-38P \\
hep-th/9805120 \\
\end{flushright}

\begin{centering}
\vspace{.1in}
{\large {\bf A Microscopic Liouville Arrow of Time }} \\
\vspace{.2in}

{\bf John Ellis$^{a}$},  
{\bf N.E. Mavromatos$^{b}$} 
and 
{\bf D.V. Nanopoulos$^{c}$}

\vspace{.15in}
 
{\bf Abstract} \\
\vspace{.1in}
\end{centering}
{\small 
We discuss the treatment of quantum-gravitational fluctuations in the
space-time background as an `environment', using the formalism for
open quantum-mechanical systems, which leads to a microscopic arrow
of time. After reviewing briefly the open-system formalism, and the
motivations for treating quantum gravity as an `environment', we
present an example from general relativity and a general framework
based on non-critical strings, with a Liouville field that we identify
with time. We illustrate this approach with calculations in the
contexts of two-dimensional models and $D$ branes. Finally, some 
prospects for observational tests of these ideas are mentioned.}

\vspace{0.15in}
\begin{flushleft}
$^{a}$ Theory Division, CERN, 1211 Geneva 23, Switzerland. \\
$^{b}$ P.P.A.R.C. Advanced Fellow, Department of Physics
(Theoretical Physics), University of Oxford, 1 Keble Road,
Oxford OX1 3NP, U.K.  \\
$^{c}$ Department of Physics, 
Texas A \& M University, College Station, TX 77843-4242, USA, \\
Astroparticle Physics Group, Houston
Advanced Research Center (HARC), The Mitchell Campus,
Woodlands, TX 77381, USA, \\
Academy of Athens, Chair of Theoretical Physics, 
Division of Natural Sciences, 28 Panepistimiou Avenue, 
Athens 10679, Greece. \\
\end{flushleft}

\vspace{0.15in}
\begin{centering}
{\it Invited Review for the} {\bf Journal of Chaos, Solitons and Fractals},
{\it Special Issue on: ``Superstrings, M, F, S, ... theory'', eds. 
C. Castro and M. S.
El Naschie.} 
\end{centering}

\end{titlepage} 

\newpage

\section{A New Time for the New Millennium?}

The nature of time remains one of the great mysteries of physics.
During the nineteenth century, the monotonic increase of entropy
incarnated in the laws of thermodynamics made manifest
a macroscopic arrow of time, but did not explain it. During the
twentieth century, time was first reinterpreted as a coordinate of
the relativistic space-time continuum, and later understood to be
distorted by gravitational fields. However, time still appeared in
both special and classical general relativity as a reversible
coordinate, although an arrow reappeared at the macroscopic level
via the expansion of the universe. Microscopically, neither did a
microscopic arrow appear in the quantum treatment of time in
Schr\"odinger's equation and its relativistic generalizations. In
quantum field theory, time reversal (T) is not necessarily an exact
symmetry, and it is known experimentally to be violated in the
neutral kaon system. Nevertheless, the combination of T with charge
conjugation (C) and parity (P) is known to be an exact symmetry of
quantum field theory, thanks to the CPT theorem~\cite{CPT} that follows
from
Lorentz invariance, locality and causality. Thus time
has no apparent arrow at the microscopic level. 

There have been many suggestions that the only arrow of time is
cosmological, and that the increase in thermodynamic entropy is due
to entanglement with the macroscopic environment. There is, however,
a minority view that there might be a microscopic arrow of time,
engendered by quantum gravity. This suggestion originates from the
formulation by Hawking~\cite{Hawk1} and Bekenstein~\cite{Bek} of 
black-hole thermodynamics,
which was first derived at the semi-classical level. It was subsequently
suggested that quantum-gravitational entropy growth might also
appear at the microscopic level~\cite{hawk}, entailing abandonment of the
$S$-matrix description of particle scattering in conventional
quantum field theory, and its replacement by a non-factorizing
superscattering operator between asymptotic in- and out-states.
As has been pointed out by Wald~\cite{ward}, any such formalism
would entail abandoning the strong form of the CPT theorem valid
in conventional quantum field theory.
If this is the case, the conventional 
microscopic quantum-mechanical time
evolution of elementary particles would need to be modified.
It has been suggested that they may need to be described as
open quantum-mechanical systems interacting with an `environment'
furnished by quantum-gravitational background fluctuations~\cite{ehns},
that
introduces a microscopic arrow of time.

Any such suggestion must be measured by the standards of the
emerging quantum theory of gravity provided by what used to be called
string theory~\cite{string}. This framework is apparently powerful enough
to
provide a consistent quantum description of microscopic 
gravitational fluctuations in the space-time vacuum background.
We have developed~\cite{emn} one approach to this problem, using the tools
of non-critical string theory with non-trivial conformal
dynamics parametrized by a Liouville field evolving
according to the renormalization group. In this approach, time
is identified as a renormalization-scale parameter provided by
the zero mode of the Liouville field, and has an arrow associated
with the orientation of the renormalization group flow.

The purpose of this article is to review (in Section 2) the basic ideas of
this formalism, which leads to a derivation of a modified quantum
Liouville equation for the density matrix, and to outline (in
Section 3) some model calculations that exemplify this general
approach. We conclude (in Section 4) by discussing the outlook
for this approach.

\section{A Brief History of Liouville Time}

\subsection{Quantum Mechanics of Open Systems: \\
$\;\;\;\;$ Density Matrix Formalism and Induced Decoherence}

A quantum-mechanical (sub)system interacting with 
some environment does not satisfy a standard Schr\"odinger equation.
The interaction with the environment 
implies that the proper description of the system is via a density matrix
$\rho \equiv Tr_{{\cal M}}|\Psi ><\Psi |$, where $|\Psi >$ is a state  
vector for the subystem, 
and $Tr_{{\cal M}}$ denotes an average over unobserved states of the
external environment. 
The temporal evolution of $\rho $ is therefore given by:
\be
   \partial _t \rho = i[\rho, H] + \nd{\delta H}(\rho )
\label{general}
\ee
where the correction $\nd{\delta H}(\rho)$ is {\it a priori} a non-linear 
function of the density matrix, describing the interaction 
of the subsystem with the environment. Linearizations of  
this interaction have been considered in the literature~\cite{lindblad,markov},
in particular in the context of quantum-gravitational
environments~\cite{ehns,hawk},
which we pursue here. If this is so, $\nd{\delta H}$ 
is an operator independent of $\rho$, which however depends on the nature 
of the environment~\cite{ehns}:
\be
   \partial _t \rho = i[\rho, H] + \nd{\delta H}\rho 
\label{ehns}
\ee
We give below a more explicit form for such linear 
environmental entanglement.

Consider, for concreteness, an open system described by a Hamiltonian 
$H$, in interaction with a quantum-mechanical environment, described 
generically by  annihilation and creation operators $B_m$, $B^\dagger_m$.
For simplicity, we assume that the environmental entanglement
conserves the total probability and energy on the average.
Such an environmental entanglement is described by the so-called 
Lindblad formalism~\cite{lindblad}  for the evolution equation 
of the density matrix of the system:
\be
{\dot \rho } \equiv \partial _t \rho
= i [\rho, H] - \sum _m
\{ B^\dagger_m B_m, \rho  \}_{+}
+ 2 \sum _m  B_m\rho  B^\dagger _m
\label{bloch}
\ee
This type of linear evolution equation for $\rho$ is used generally
to describe Markov processes in open quantum-mechanical
systems~\cite{markov}. Recalling that
the density matrix may be expressed in terms 
of the state vector $|\Psi >$ via 
\be
  \rho = {\rm Tr}_{\cal M} |\Psi > < \Psi |
\label{neumanform}
\ee
with the trace being taken over an ensemble of theories ${\cal M}$,
it is possible to derive a
time-evolution equation for the state vector $|\Psi >$, 
which is of stochastic Ito form~\cite{gisin}:
\bea
&~&|d\Psi > =-\frac{i}{\hbar} H |\Psi >
+ \sum _m (<B^\dagger _m >_{\Psi} B_m - \frac{1}{2}
B^\dagger _m B_m - \nn \\
-&~&\frac{1}{2} <B^\dagger _m>_{\Psi}
<B_m>_{\Psi })~|\Psi > dt +
\sum _m
(B_m - <B_m>_{\Psi})~|\Psi > d\xi _m
\label{ito}
\eea
where 
$<\dots>_{\Psi}$ denote averages with respect to the
state vector $|\Psi >$, and $d\xi _m$ are complex differential
random matrices, associated with white-noise Wiener or
Brownian processes. 

Within the stochastic framework (\ref{ito}), it
can be shown, under assumptions on the Hamiltonian operator
that are not too restrictive,
that $|\Psi >$ will always become localized in some
state-space channel $k$, as a result of
environmental entanglement.
To prove this formally, one may construct a quantity that serves as a
measure of the delocalization
of the state vector, and examine its temporal
evolution. An example is the
quantum dispersion entropy~\cite{gisin}:
\be
{\cal K} \equiv -\sum _k <P_k>_{\Psi}log~<P_k>_{\Psi}
\label{qde}
\ee
where $P_k$ is a projection operator onto the state $k$ of the system. 
This entropy has been shown to decrease in situations where
(\ref{ito}) applies, under some
assumptions about the commutativity of the
Hamiltonian of the system with $P_k$, which implies
that $H$ can always be written in a block-diagonal form.

Analyses in the above framework have yielded
important results
concerning the passage from the quantum to the classical 
world~\cite{ehns,zurek,gisin,emn}. 
In particular, in~\cite{zurek} it was suggested
that classical states may appear as a result of quantum
decoherence, depending on the nature of the environment~\cite{albrecht}. 
Such states are minimum entropy/uncertainty states
whose shape 
is retained during evolution. 
Such `almost classical' states are termed 
`pointer states' by Zurek~\cite{zurek}.  
Such pointer states emerge, as a result of decoherence, 
when the localization process of the state vector has stopped
at a stage, not where 
it is complete, but so that the resulting minimum-entropy
state is least susceptible to the effects of the environment. 

We close this section by recalling a
recent experiment~\cite{brune} in which
a mesoscopic `Schr\"odinger's cat' was constructed,
and the associated quantum decoherence was observed, for the first time 
in experimental physics. 
The experiment uses electromagnetic cavities of the type familiar from 
non-demolition experiments in quantum optics~\cite{haroche,qnd}. 
The `cat' is constructed in two stages:
first it involves an interaction of the two-state atom with the 
cavity field, which results in a coherent state of the 
combined `atom + meter' system, and then dissipation is induced 
by coupling the cavity (interpreted as a measuring apparatus) to the
environment, thereby inducing decoherence
in the `atom + meter' system. The important point is
that the more macroscopic is the cavity mode, i.e., the higher 
the number of oscillator quanta, the shorter is the decoherence
time. This is exactly what was to be expected 
from the general theory~\cite{zurek,ehns,emohn,emn}. 

\subsection{Quantum Gravity as an Environment: \\
$\;\;\;\;$ A Microcopic Arrow of Time?}

We now examine whether the above approach may be applicable to
quantum gravity. Here the chief problem is the present lack of a
consistent mathematical formalism for quantum
fluctuations of the gravitational field. 
The appearance of singular space-time configurations, 
such as black holes, is one of the problems that 
prevent us from constructing a satisfactory local 
quantum field theory for the gravitational field. 
Indeed, the presence of microscopic event horizons, as around black
holes of Planckian size, is incompatible with 
the conventional unitary temporal evolution 
of quantum mechanics. 

To understand this, consider the possibility of an 
asymptotic initial pure quantum state of matter at $t \rightarrow -
\infty$, characterized by two sets 
of quantum numbers $\{ A \}$ and $\{ B \}$. 
Suppose that a measurement is made in the asymptotic future, $t
\rightarrow \infty$, 
by an observer located at spatial 
infinity from a black hole which is formed, either by collapsing matter or 
by quantum fluctuations, at the time $t=0$. 
Assume further that the quantum numbers $\{ B \}$ are absorbed by the
black hole. 
Due to quantum fluctuations 
the black hole will emit Hawking radiation~\cite{hawk}
and eventually evaporate, apparently without any memory of the quantum
numbers $\{ B \}$. Hence an
asymptotic observer at future spatial infinity,
who sees around him/herself a macroscopically-flat space 
time, has to average over this set of quantum numbers,
corresponding to an average over the unobservable states of the black
hole. This implies that he/she is forced to use a density
matrix $\rho_{out}$ to describe the asymptotic state, 
which is no longer a pure state, but rather a statistical mixture. 
This in turn requires non-unitary evolution, since pure states 
evolve into mixed ones. 

This incompatibility with conventional quantum mechanics
suggests that the quantum-mechanical system becomes {\it open} in the
presence
of singular quantum-gravitational configurations or fluctuations.
Quantum gravity acts as an environment, whose averaging prevents
factorization 
of the scattering matrix. Hawking~\cite{hawk}  
has argued that the relevant well-defined concept is that of
the superscattering matrix $\nd{S}$ introduced to
generalize the scattering matrix in the case of statistical mixtures of 
asymptotic states:
\be
     \rho _{out} = \nd{S} \rho _{ij}~, \qquad \nd{S} \ne SS^\dagger  
\label{superscattering} 
\ee
where $S \sim e^{iHt}$ is a conventional unitary Heisenberg scattering
matrix.

Non-factorizability of the superscattering would
imply generically some breaking
of the strong form of CPT symmetry, as shown by Wald~\cite{ward}.
Moreover, there is entropy production, in agreement with the open
nature of the observable subsystem. Passing from the asymptotic
description (\ref{superscattering}) to finite-time evolution,
one expects a modified Liouville equation for the density
matrix of the form (\ref{ehns}), (\ref{bloch}), where the `environment'
operators now represent
quantum-gravitational coherent states~\cite{ehns,emohn}.

An explicit toy model for the interaction of 
low-energy propagating matter 
with a dissipative quantum-gravity 
environment consisting of 
virtual wormholes~\cite{wormhole} was studied in~\cite{emohn}
from a phenomenological
viewpoint. Coleman argued that the wormhole state 
was likely to be coherent, and used this argument
to support the the vanishing of the 
cosmological constant. 
However, this coherence assumption was questioned later, and
our subsequent studies of the nature of the space-time 
foam in quantum gravity and string theory
suggest that this is not the case. 
However, one can still construct a model interaction Hamiltonian 
between operators describing the low-energy probe $O_P$ and 
the wormhole state $|a>$~\cite{wormhole}:
\be
   H_I \propto O_P (a^\dagger + a ) 
\label{wormholes}
\ee
where $a^\dagger,a$ are creation and anihilation operators
for wormhole states. In the example of~\cite{emohn},
$O_P$ was taken to be a four-fermion effective interaction
\be
  O_P \propto {\cal O}(\frac{1}{m_P^2}) {\overline \psi}_1
\gamma^\mu \psi_1 {\overline \psi}_2 \gamma_\mu \psi_2
\label{fourfermi}
\ee
A low-energy observer has to average out the
unobservable wormhole 
effects, with the result that 
the low-energy probe $P$ becomes an open system. 
The simple case of a Gaussian 
distribution for the wormhole configurations 
was assumed in~\cite{emohn}, and the time scale of the 
induced decoherence
of the low-energy probe $P$ was estimated,
using the phenomenological 
equation for the density matrix suggested in~\cite{ehns}, which was
characterized by probability 
and energy conservation of the probe.
In the framework of the previous section, 
this coupling may be considered as providing
only phase damping for the 
atom. The situation resembles that of the atom + cavity 
system of~\cite{brune}, with the essential difference that the
effect is not due to quantum 
electrodynamics, but to quantum-gravitational 
intreractions. The r\^ole of the cavity is played by
the microscopic space-time foam~\cite{hawk,ehns}. 

As was shown in~\cite{emohn}, the wormhole-probe coupling
is enhanced for large numbers of atoms. Consequently,
the decoherence of off-diagonal elements of 
the density matrix $\rho (x,x')$, in a `pointer' basis 
$|x>$, where $x$ is the center-of-mass location in space time 
of a system of $N$ particles, is of the form~\cite{emohn}: 
\be
\rho (X',X, t) \sim \rho_0 (X',X, t) {\rm exp}[-ND(X'-X)^2t]
\label{decohmoha}
\ee
where $D$ represents the coupling of a single 
particle with a single coherent mode of the wormhole state, and is 
estimated to be of order $D \sim m^6/M_P^3$ for a particle of mass $m$
in a four-dimensional space time, where $M_P$ is the Planck mass. 
A uniform density of 
wormholes of the order of one per Planck volume in space time
was assumed in making the estimate (\ref{decohmoha}),
and all other interactions
of the microscopic particles among themselves have been ignored. 
One can readily see from (\ref{decohmoha}) 
the characteristic feature that the decoherence rate 
is proportional to the square of the distance between 
the pointer states~\cite{zurek}, which is a generic
feature of Markov-type decoherence~\cite{markov,milburn}.

The above description by no means 
constitutes a microscopic description of decoherence in quantum gravity.
Because the underlying theory is still unknown, one cannot
describe the pertinent `environment' in a detailed way. However, 
in subsequent sections of this review we present a formalism which
may provide a way to formulate the task. We preface our discussion with
a suggestive example from general relativity, and then move to 
an analysis in the context of 
a non-critical Liouville string approach. 

\subsection{Time as a Renormalization-Group Scale: \\
$\;\;\;\;$ A Suggestive Example from General Relativity}

Aspects of the above intuitive arguments are
illustrated in a study\cite{MW} of
a scalar field of mass $\mu$
coupled to an Einstein-Yang-Mills (EYM) black hole
in four dimensions. This study was purely in the context of
quantum gravity, with no string degrees of freedom, but
features gauge hair and entanglement entropy.
The renormalizability of the EYM field theory
enables the partition function and entropy of the
four-dimensional black hole to be calculated
with the inclusion of quantum corrections~\cite{MW}.

Some of the
quantum corrections to the entropy
can be absorbed into the semi-classical Hawking-Bekenstein
entropy $S = A / (4 G_N)$, where A is the area of the horizon, via a
renormalization of Newton's
constant $G_{N}:~G_{N_0} \rightarrow G_{N_{ren}}$. However,
not all the quantum corrections can
be absorbed into a bare parameter in this way,
pointing to a new
effect beyond the reach of the conventional renormalization
programme. Formally, it may be absorbed
into a quantum `horizon area', which
depends logarithmically on
an ultraviolet cutoff $\epsilon$, and induces similar
dependences in the partition function $Z$ and the entropy $S$:
\be
F \equiv - {\rm log} Z, \; \; \;S \sim {\rm log} ({\epsilon \over r_h})
\label{logarithms}
\ee
where $r_h$ is the semi-classical horizon radius.
Since Hawking radiation causes $r_h$ to vary in time,
we~\cite{emnw} identify 
${\rm log}\left(\epsilon/r_h\right)$ with the target time $t$,
fixing the sign by 
consistency with Hawking's quantum evaporation of the black hole~\cite{hawk}.

As shown in more detail in~\cite{emnw}, this induced 
time-evolution of the density matrix of the scalar field in the 
EYM black hole background cannot be described by a conventional
Hamiltonian $H$. It requires an open
quantum-mechanical evolution equation of the form (\ref{ehns}), with
\bea
\nd{\delta H} & = & \beta \left(- \frac {\partial H}{\partial t}
+\frac {\partial F}{\partial t} \right) \\
& = & {\cal O}({\mu^2 \over M_P}) + {\cal O}\left(\hbox{log} ({\epsilon
\over
r_h})\right)^{-1} + \dots ,
\label{Hslash}
\eea
and the entropy of the scalar field $\phi$ increases with time:
\be
\frac {\partial S}{\partial t} =
{\cal O} ({\mu^2 \over M_P}) + {\cal O} ({r_h \over \epsilon}) + \dots
\label{Sdot}
\ee
The fact that the scalar field $\phi$ state 
becomes more mixed as a result of
the $\nd{\delta H}$ term (\ref{Hslash}) in the quantum
Liouville equation (\ref{ehns}) is
due to a change in the entanglement of the external $\phi$ field
with the unmeasurable modes interior to the black hole.
The amount of this entanglement depends on the quantum numbers
of the EYM background. In particular, the more gauge
hair it has, the smaller the $\epsilon$ dependence in 
the partition function~\cite{emnw},  
and hence the slower the rate of information
loss in (\ref{Sdot}).

The quadratic dependence of $\nd{\delta H}$ on the
scalar mass $\mu$, divided in order of magnitude by just one
power of $M_P$, is the largest that could be expected for any
such modification of the quantum Liouville equation:
{\it a priori}, it could have been suppressed by one or more
additional powers of $\mu / M_P$, or an exponential, or
even absent all together. We are not in a position
to estimate the coefficient of this $\mu^2/M_P$ term.
Nor, indeed, can we be sure that such a parametric dependence
would persist in a complete quantum theory of gravity. 
However, in Section 3 we review some explicit calculations
within the Liouville string approach to quantum gravity, 
where we show that such a quadratic dependence indeed appears 
as a possibility~\cite{gang,ekmnw}. 

\subsection{Liouville-String Approach to Quantum Gravity: \\
$\;\;\;\;$ Time as a World-Sheet Renormalization-Group Scale}

This is not the place to review the great promise that
string theory~\cite{string}, now in its non-perturbative incarnation as
$M$ theory, holds as a possible consistent quantum theory
of gravity. Instead, we focus on the heterodox aspects of
our Liouville string approach~\cite{ddk,mm}. Most discussions of string
theory concentrate on classical backgrounds, which are
described by conformal field theories on the string world sheet.
In this case, one has a decoupling of
the Liouville mode $\phi$ that scales the 
world-sheet metric:
\be
\gamma_{\alpha\beta}=e^{\varphi}{\hat \gamma}_{\alpha\beta}
\equiv e^{\phi/Q}{\hat \gamma}_{\alpha\beta}
\label{Lscale}
\ee
where ${\hat \gamma}$ a some suitable fiducial metric
on the world sheet, and $Q$ is defined below, and the
Liouville dynamics is trivial as a result of conformal symmetry.
Instead, we take the point of view that in order to discuss
quantum fluctuations in the string background or transitions
between generic different string vacua, one needs a broader
framework than conformal field theories on the world sheet.

We therefore
consider a generic conformal field theory action $S[g^*]$ perturbed by
non-conformal deformations $\int d^2z g^iV_i $, whose couplings 
have non-trivial world-sheet renormalization-group $\beta$-functions:
\be
\beta_i=(h_i-2)(g^i-g^{*i})
+ c^i_{jk}(g^j-g^{*j})(g^k-g^{*k}) + \dots ,
\label{betaf}
\ee 
where the $c^i_{jk}$ are operator product expansion 
(OPE) coefficients defined in the normal way. Coupling this theory to 
two-dimensional quantum gravity restores conformal invariance at the quantum 
level, with the Liouville field acting as a dynamical 
local scale on the world-sheet, that makes
the gravitationally-dressed operators $[V_i]_\phi$ 
exactly marginal. The corresponding 
gravitationally-dressed conformal theory is:
\be
S_{L-m} = S[g^*] + \frac{1}{4\pi\alpha '}
\int d^2 z \{\partial _\alpha \phi \partial ^\alpha \phi
- QR^{(2)} + \lambda ^i(\phi ) V_i \}
\label{C5}
\ee
where~\cite{ddk,polkle}:
\be
\lambda ^i(\phi ) =g^i e^{\alpha _i \phi }
+ \frac{\pi}{Q \pm 2\alpha _i } c^i_{jk} g^jg^k
\phi e^{\alpha _i \phi } + \dots~~;~~\alpha _i = -\frac{Q}{2} +
\sqrt{\frac{Q^2}{4} - (h_i - 2)}
\label{C6}
\ee
Within this general framework, we focus
on the operators $V_i$ that are $(1,1)$ but not exactly marginal,
i.e., have $h_i=2$ but $c^i_{jk} \ne 0$. Their couplings obey 
$\frac{d}{d \tau}\lambda^i(\phi)=\beta^i$, where 
$\tau=-\frac{1}{\alpha Q}{\rm log}A:~A\equiv \int d^2z \sqrt{{\hat
\gamma}}
e^{\alpha\phi(z,{\bar z})}$ is the world-sheet area, and 
$\alpha = -\frac{Q}{2}+\frac{1}{2}\sqrt{Q^2+8}$ with:
\be
Q=\sqrt{\frac{|25-C[g,\phi]|}{3g_s^\chi} + \frac{1}{2}\beta^iG_{ij}\beta^j}
\label{31/2}
\ee
Here $C[g,\phi]$ is the Zamolodchikov $C$  function~\cite{zam},
which is given according to
the $C$ theorem~\cite{zam} by: 
\be  
  C[g, \phi]=c^*-g_s^\chi \int_{\phi_*}^{\phi}d\phi'\beta^iG_{ij}\beta^j
\label{ctheorem}
\ee 
where $*$
denotes a fixed point of the world-sheet flow,
$g_s$ is the string coupling
and $\chi$ is the Euler characteristic of the world-sheet manifold.
The other part of $Q$ is due to the local character of the 
renormalization-group scale~\cite{emn}, with $G_{ij}$ related to divergences 
of $<V_iV_j>$ and hence the Zamolodchikov metric~\cite{zam} in the
space of possible theories. 

We interpret such
non-critical strings~\cite{aben,emn} 
as effective string models that describe the dynamics of modes
observable in {\it local} scattering experiments. 
It is known that generic string backgrounds, such as black holes,
also have unobservable {\it delocalized} modes which 
combine with local modes to provide exactly marginal
deformations in the presence of singular space-time fluctuations.
The corresponding local modes are not exactly
marginal, and should be described by the formalism of the
previous paragraph.
The couplings of such local modes to delocalized modes
lead to quantum entanglement and the possibility of information loss
and entropy growth.

The zero mode of the Liouville field in such a string 
theory may be identified~\cite{aben,emn,kogan} 
with a target-time variable, since its kinetic term
has negative metric, as seen in (\ref{C5}) and as contrasted with
the kinetic terms for flat space coordinates in the fixed-point
action:
\be
 S[g^*]=-\frac{1}{4\pi \alpha '} \int d^2z \partial_\alpha X^i \partial ^\alpha X^i 
\label{flat} 
\ee
The opposite sign between (\ref{C5}) and (\ref{flat}) arises because the
effective non-critical string theory is
{\it supercritical}, i.e.,
its matter central charge is greater than the critical (conformal) value,
which is 25 for the bosonic strings discussed here~\cite{aben,ddk,mm}.
This identification of
time with the Liouville field entails a modification of the
conventional Hamiltonian time evolution, reflecting
the above-mentioned quantum entanglement,
which results in an irreversible 
temporal evolution in target space,
with decoherence and associated entropy production, as we now review.

The effective density matrix for the local modes is:
\begin{equation}
\tilde \rho (local, t) \, = \, \int d(delocal)\rho (local, delocal)
\label{intout}
\end{equation}
where ${\rho}$ denotes the full density matrix, and
the $delocal$ states play a role analogous to those of the
unseen states inside the black-hole horizon
in the arguments of~\cite{hawk} and the previous subsections.
The integration over $delocal$ in (\ref{intout})
ensures that the reduced density matrix $\tilde \rho$ is mixed in
general, even if the full $\rho (local, delocal)$ is pure. We have
shown that $\tilde \rho$ obeys a modified quantum Liouville
equation of the form (\ref{ehns})~\cite{emn,kogan} with a
Hamiltonian $H$ and
\begin{equation}
\nd{\delta H} = -i \Sigma_{i,j} \beta^i G_{ij}[\, \, , g^j]
\label{stringmodliou}
\end{equation}
where the indices
$(i,j)$ label all possible microscopically-distinct string
background states with coordinate parameters $g^i$, and $G_{ij}$
is a metric in the space of such possible backgrounds~\cite{zam}. 
Since these are not conformally invariant once one integrates out
the $delocal$ degrees of freedom, the 
corresponding renormalization functions $\beta^i$ are
non-trivial, reflecting the
back reaction of the light particles on the
background metric. We note further that the background 
fields $g^i$ must be quantized, as a result of the summation
over world-sheet 
topologies in the Liouville string~\cite{emn}. 
In the toy context of two-dimensional black-hole space times,
the $delocal$ degrees of freedom are topological higher-level 
discrete string modes, which represent infinite-dimensional stringy gauge 
symmetries described mathematically by $W_\infty$ 
algebras~\cite{EMNcount,emn,chaudhuri}. 

There are general properties of the Liouville system that follow
from the renormalizability 
of the world-sheet $\sigma$-model theory~\cite{emn}.
These include {\it energy conservation} on the average, 
and 
{\it probability } conservation. The nature of
the renormalization group on the two-dimensional
world-sheet~\cite{zam} entails {\it monotonic} entropy
increase,
$\partial _t S \propto \beta^i G_{ij} \beta^j \ge 0$, 
and hence a {\it microscopic} arrow of time.
The maximum magnitude of effect that we can imagine is
\begin{equation}
\nd{\delta H} \simeq H^2 / M_P
\label{order}
\end{equation}
which would be around $10^{-19} ... 10^{-20}$ GeV for a typical 
low-energy probe.
A contribution to the evolution rate equation
(\ref{stringmodliou}) of this order of magnitude would arise if there
were some Planck-scale interaction contributing an amplitude
$A \simeq 1/M_P^2$ and hence a rate $R \simeq 1/M_P^4$, to be
multiplied by a density of singular microscopic backgrounds
$n \simeq L_P^{-3} \simeq M_P^3$, yielding the
overall factor of $ \simeq 1/M_P$ shown in 
(\ref{order})~\cite{emohn}. As we have seen above, such an 
estimate was found in a pilot study of a scalar field in a 
four-dimensional black-hole background~\cite{emnw}, and
such an estimate is also found in a  
Liouville-string representation of Dirichlet membranes
($D$ branes)~\cite{emnd},
as we shall discuss in section 3.

Similar conclusions are reached in a study of
asymptotic scattering, since it is impossible to
define factorizable $S$-matrix elements in Liouville string theory, 
once we interpet the Liouville field as time.  
To see this, we first note that 
correlation functions in Liouville theory may be written in the form
\be
A_N \equiv <V_{i_1} \dots V_{i_N} >_\mu = \Gamma (-s) \mu ^s
<(\int d^2z \sqrt{{\hat \gamma }}e^{\alpha \phi })^s {\tilde
V}_{i_1} \dots {\tilde V}_{i_N} >_{\mu =0}
\label{C12}
\ee
where the ${\tilde V}_i$ 
have the Liouville zero mode removed, $\mu$ is a 
scale related to the world-sheet cosmological constant, 
and $s$ is the sum of the anomalous dimensions of the 
$V_i~:~s=-\sum _{i=1}^{N} \frac{\alpha _i}{\alpha } - \frac{Q}{\alpha}$.
As it stands, (\ref{C12}) 
is ill defined for $s=n^+ \in Z^+$, because of the 
$\Gamma (-s)$ factor~\cite{emndollar}. 
To regularize this factor, we use the 
integral representation~\cite{kogan2,emn} 
$\Gamma (-s)=\int dA e^{-A} A^{-s-1}$, 
where $A$ is the covariant area of the world sheet, 
and analytically continue to the contour shown in 
Fig. 1. Intepreting the Liouville field $\phi$ 
as time~\cite{emn}: $t \propto {\rm log}A$, we interpret 
the contour of Fig. 1 as representing evolution 
in both directions of time between fixed points of the 
renormalization group: $ {\rm Infrared} ~ {\rm fixed} ~ 
{\rm point}  \rightarrow {\rm  Ultraviolet} ~ {\rm fixed}
~{\rm point} \rightarrow
 {\rm Infrared} ~ {\rm fixed} ~ {\rm point}$.

\begin{figure}
\hglue3.0cm
\epsfig{figure=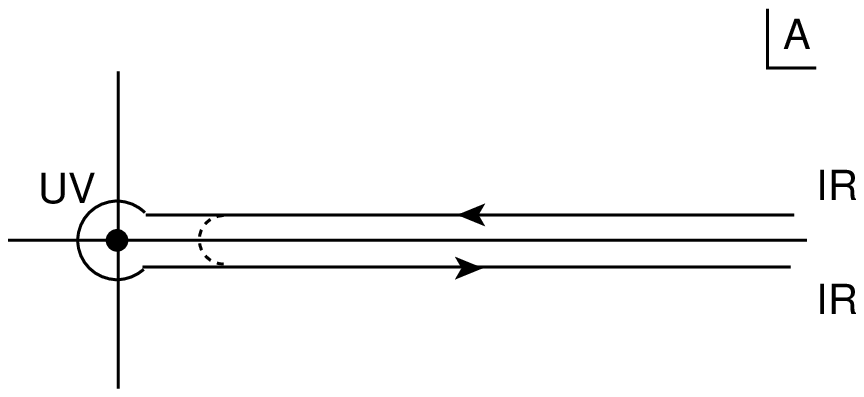}
\caption{ {\it The solid line is the       
the Saalschutz contour in the complex
area ($A$) plane, which is used to 
continue analytically the prefactor 
$\Gamma (-s)$ for $ s \in Z^+$; 
it has been used in
conventional quantum field theory to relate dimensional
regularization to the Bogoliubov-Parasiuk-Hepp-Zimmermann
renormalization method. The dashed line denotes the 
regularized contour, which avoids the ultraviolet 
fixed point $A \rightarrow 0$, which is used in the Closed Time-like Path 
formalism.}}
\label{saaltz}
\end{figure}

Within this approach, it is not difficult to see that conventional 
$S$-matrix elements are 
in general ill-defined in Liouville-string theory, and that
scattering 
must be described by a non-factorizable $\nd{S}$-matrix. 
Decomposing the Liouville field in an orthonormal 
mode sum: 
$\phi (z, {\bar z}) = \sum _{n} c_n \phi _n  = c_0 \phi _0
 + \sum _{n \ne 0} \phi _n$, 
where $   \nabla ^2 \phi _n = -\epsilon_n \phi _n~:~n=0, 1,2, \dots$,
we separate the zero mode 
$\phi _0 \propto A^{-\frac{1}{2}}$
with $\epsilon _0 =0$. One may now consider ${\tilde A}_N$, the
correlation function with $\phi_0$ subtracted, and its
behaviour under an infinitesimal Weyl transformation  
$\gamma (x,y) \rightarrow \gamma (x,y) ( 1 - \sigma (x, y))$.
We have found that the correlator ${\tilde A}_N$
transforms as follows~\cite{emnd}:
\bea
\delta {\tilde A}_N \propto &~&
\sum _i h_i \sigma (z_i ) + \frac{Q^2}{16 \pi }
\int d^2x \sqrt{{\hat \gamma }} {\hat R} \sigma (x) +    \nn \\
&~& {\cal O}\left( s \over {\hat A} \right)
\label{dollar}
\eea
where the hat notation denotes transformed quantities.
We see explicitly that (\ref{dollar}) contains non-covariant 
terms $\propto A^{-1}$ 
if the sum of the anomalous dimensions
$s \ne 0$. Thus the generic correlation 
function $A_N$ does not have a well-defined 
limit as $A \rightarrow 0$.

In our approach to string time we identify~\cite{emn} 
the target time as $t=\phi_0=-{\rm log}A$, 
where $\phi_0$ is the world-sheet zero mode of the Liouville field.
The normalization follows from
a consequence 
of the canonical form of the kinetic term for the Liouville field $\phi$ 
in the Liouville $\sigma$ model~\cite{aben,emn}. 
The opposite flow of the target time, as compared to that of the 
Liouville mode, is, on the other hand, a consequence
of the `bounce' picture~\cite{kogan2,emn} for Liouville flow of Fig. 1.
This identification implies that, as a result of the above-mentioned
singular behaviour in the ultraviolet limit $A \rightarrow 0$,
the correlator ${\tilde A}_N$
cannot be interpreted as an $S$-matrix element, 
whenever there is a departure from criticality $s \ne 0$.

When one integrates over the Saalschultz contour in Fig. 1, the integration
around the simple pole at $A=0$ yields an imaginary part~\cite{kogan2,emn},
associated with the instability of the Liouville vacuum. We note, on the  
other hand, that the integral around the dashed contour
shown in Fig. 1, which does not encircle the pole at $A=0$, is well defined.
This can be intepreted as a well-defined $\nd{S}$-matrix element,
which is not, however, factorisable into a product of 
$S-$ and $S^\dagger -$matrix elements, due to the 
$t$ dependence acquired after the identification 
$t=-{\rm log}A$~\footnote{This formalism is similar to the 
Closed-Time-Path (CTP) formalism used in non-equilibrium 
quantum field theories~\cite{ctp}.}.

\section{Play Times}

In this section we shall present briefly some 
examples and applications
of the above Liouville string formalism. Although the 
examples considered are simple, we believe
these toy cases capture important qualitative feature
of realistic string gravity situations.
 
\subsection{Two-Dimensional String Black Holes}

As a first illustration of our approach to non-critical string theory,
we discuss the two-dimensional black-hole~\cite{witt}.
This toy laboratory gives insight into the nature
of time in string theory and contributes to the physical effects
mentioned in other sections. Its action imay be written
in the form~\cite{witt}
\be
   S_0=\frac{k}{2\pi} \int d^2z [\partial r {\overline \partial } r
- {\rm tanh}^2 r \partial t {\overline \partial } t] + \frac{1}{8\pi}
\int d^2 z R^{(2)} \Phi (r)
\label{action3}
\ee
where $r$ is a space-like coordinate and $t$ is time-like,
$R^{(2)}$ is the scalar curvature, and $\Phi$ is the dilaton field. The
customary interpretation of (\ref{action}) is as a string model with
$c$ = 1 matter, represented by the $t$ field, interacting with a
Liouville mode, represented by the $r$ field, which has $c  < 1$ and
is correspondingly space-like~\cite{aben}.
As an illustration of the
approach outlined in the previous section, however, we re-interpret~\cite{emn}
(\ref{action}) as a fixed point of the renormalization group
flow in the local scale variable $t$. In our interpretation, the
`matter' sector is defined by the spatial coordinate $r$, and has
central charge $c_m$ = 25 when $k  = 9/4$~\cite{witt}. Thus the model
(\ref{action3}) describes a critical string in a dilaton/graviton
background. The fact that this is static, i.e., independent of $t$,
reflects the fact that one is at a fixed point of the renormalization
group flow \cite{emn}.

We now outline how one can use the machinery
of the renormalization group in curved space,
with $t$ introduced as a local Liouville renormalization
scale on the world sheet, to derive the model (\ref{action}): a detailed
technical description is given in~\cite{emn}.
There are two contributions
to the kinetic term for $t$ in this approach, one associated with
the Jacobian of the path integration over the world-sheet metrics, and
the other with fluctuations in the background metric.

To exhibit the former, we first choose the conformal gauge
$\gamma _{\alpha \beta}=e^{\varphi}
{\hat \gamma}_{\alpha\beta}$~\cite{ddk,mm},
where $\varphi$ represents the Liouville mode, as in (\ref{Lscale}).
An explicit computation~\cite{mm} of the Jacobian using heat-kernel
regularization yields
\be
-\frac{1}{2\pi}
\partial_\alpha \varphi \partial^\alpha \varphi + \dots
\label{liouvillekin}
\ee
where the dots represent terms that are not essential for
this argument, and an appropriate overall normalization has been chosen. 
This procedure
reproduces the critical string results of~\cite{witt}
when one identifies
the Liouville field $\varphi$ with $\sqrt{k} t$. Equation
(\ref{liouvillekin})
contains a negative (time-like) contribution to the kinetic term
for the Liouville (time) field, but this is not the only such
contribution. Quantum fluctuations in the background metric yield,
after renormalization, a term of the form~\cite{emn}
\be
\frac{k}{8\pi}R \partial_\alpha t
\partial^\alpha t
\label{scalarcurv}
\ee
where $R$ is the scalar curvature in target space. 

In the case of the Minkowski black-hole model~\cite{witt},
the scalar curvature is
\be
R=\frac{4}{{\rm cosh}^2r}=4-4 {\rm tanh}^2r,
\label{curba}
\ee
Substituting (\ref{curba}) into equation (\ref{scalarcurv})
we obtain the
second contribution to the kinetic term for the Liouville field $\phi$.
Combining it with the world-sheet metric Jacobian term in
(\ref{liouvillekin})
we recover precisely the action
(\ref{action3})
for the two-dimensional black hole. 

This is a non-trivial check of
our identification of time with the Liouville field. The same model
can also be used to exemplify the fact that exactly-marginal
operators may in general contain combinations of local and delocalized
modes. One of the exactly-marginal operators of 
Euclidean version of the two-dimensional
black hole~\cite{witt} takes the form~\cite{chaudhuri}
\be
L_0^1{\overline L}_0^1 \propto
{\cal F}^{c-c}_{\frac{1}{2},0,0} + i(\psi^{++}-\psi^{--}) + \dots
\label{margintax}
\ee
where the $\psi$ denote higher-level operators,
and the first term in (\ref{margintax}) is a
`tachyon' operator given by
\be
{\cal F} ^{c-c}_{\frac{1}{2},0,0}(r)
=\frac{1}{{\rm cosh}r}
F(\frac{1}{2},\frac{1}{2} ; 1, {\rm tanh}^2r )
\label{tachyon}
\ee
The higher-level string modes
cannot be detected in local scattering
experiments, because of their delocalized character.
An `experimentalist' therefore sees necessarily a
truncated matter theory, where the only deformation  is the
tachyon ${\cal F}_{-\frac{1}{2},0}^c $, which is a (1,1) operator,
but is not
exactly marginal~\cite{emn}. When truncated to such local modes, the model
is non-critical, and the general analysis of the previous section
applies.

\subsection{The Arrow of Liouville Time in Two-Dimensional String
Universes}

As another non-trivial application of the Liouville approach to time, 
we consider the dilaton-matter black-hole solution found
in~\cite{tachyon}. 
In two dimensions, the action of the dilaton-tachyon system coupled to
gravity is
 \begin{equation}
 S=\frac{1}{2\pi}
 \int d^2x\sqrt{-g}
 \{e^{-2\Phi}
 [R+4(\nabla\Phi)^2-(\nabla T)^2
 -V(T)+4\lambda^2 ] \}
 \label{action}
 \end{equation}
where $4\lambda ^2$ is the two-dimensional cosmological constant,
which is related to the central charge $c$ of the corresponding
world-sheet $\sigma$ model via~\cite{witt}: 
\be
    \lambda^2=\frac{k}{3}(c-26) 
\label{lamb}
\ee
where $k >2$ is the level parameter 
of the Wess-Zumino conformal field theory that has as a 
conformal solution 
the target-space two-dimensional theory (\ref{action}), 
which is in turn related to the central charge by~\cite{witt} 
$c= (3k / k-2)-1$. The quantity 
$V(T)$ is the tachyon potential, which is ambiguous
in string theory~\cite{Banks}. 
The only unambiguous term is 
the quadratic term for the tachyon field, which 
is also all that we need for our analysis.
One solution of the equations of motion derived from this action 
is the static black hole of~\cite{witt}, which from a string theory point
of view is an equilibrium (conformal) fixed point 
solution in the space of 
string theories. A non-equlibrium exact solution appears~\cite{tachyon}
when one considers collapsing tachyonic (massless) matter,
$T= T(x^+)$, where $x^+$ is a light-cone coordinate. 
The solution is characterised 
by an initial singularity at $x^+=0$
and the absence of a white hole. 

Its stability has been analyzed 
in~\cite{gang}, whose results we now summarize.
The metric element for the $\lambda = 0$ case
can be written in the form
\be
   ds^2= e^{2\tau}\left( -\xi^2 d\tau ^2 + d\xi ^2 \right) 
\label{rindler} 
\ee
where $\xi \equiv e^{r/2}$ and $\tau \equiv t/2$. we see from
(\ref{rindler}) that this expanding space is conformally equivalent to a
two-dimensional 
Rindler space with constant acceleration~\cite{birrel}.
An analysis of Bogolubov coefficients has shown that this
$\lambda =0$ model
is thermodynamically {\it stable} in the sense of
general relativity, with no particle production.
This is because of a cancellation of
the effects of the spatial expansion visible in (\ref{rindler})
and the Rindlerlike acceleration~\cite{unruh},
as far as particle production is concerned.
On the other hand, the $\lambda \ne 0$ curved space time,
formed by collapsing matter $T(x^+)$, 
corresponds to 
time-dependent black-hole solution~\cite{tachyon}, and was found
to exhibit nonthermal particle creation, and to be
thermodynamically {\it unstable}.

It is tempting to guess that the equilibrium case $\lambda = 0$
constitutes the end-point of the time evolution
of the unstable $\lambda^2 \ne 0$ solutions.
As we have argued in \cite{gang},
this point of view is supported by a
Liouville-string interpretation of the space-time metric
(\ref{rindler}), when one identifies $t$ above with the 
Liouville zero mode. From this point of view, 
the rate of change of the cosmological constant 
during the non-equilibrium phase is
obtained from the Zamolodchikov $C$ theorem~\cite{zam}, appropriately
extended 
to Liouville strings, once one identifies the Liouville 
field with a local renormalization scale on the world-sheet~\cite{emn}:
\be
\frac{ \partial}{\partial t} \lambda ^2  \sim - \sum _{i,j}~\beta ^i <V_iV_j> \beta^j 
\label{ctheorem2}
\ee
where the $V_i$ are the 
vertex operators corresponding to deformations $g^i$, and the summation
over $i, j$ also includes
spatial integrations $\int dr \int dr ' $.
The two-point correlators $<V_i V_j>$ constitute 
a metric in theory space~\cite{zam}, which is positive definite 
for unitary world-sheet theories, 
implying an irreversible time flow~\cite{emn}.

Within this approach, 
it is immediate to deduce from (\ref{ctheorem}) 
the time dependence of $\lambda (t)$.
In the case 
of small $\lambda$ and weak matter fields $T$, the leading-order 
effect is associated with the graviton $\beta$ functions.
To ${\cal O}(\alpha ')$, the latter are  
proportional to $R G_{MN}$, where $R$ is the scalar curvature for the
generic 
case $\lambda \ne 0$ of the time-dependent black hole solution 
of \cite{tachyon}, 
computed  in~\cite{gang}.
Using the 
the initial
condition $\lambda ^2 (0)=\lambda _0 ^2 << 1$, and 
the
fact that, to this order, the
Zamolodchikov  metric $<V_iV_j>$ is proportional to 
$\delta _{ij}\delta (r - r') $, one finds:
\be
       \lambda ^2 (t) = \frac{\lambda _0^2 }{1 + \lambda _0^2 A (e^t - 1)}
\label{solution}
\ee
where 
\be
A=\left (4C_1^2 a^2 
\int _0^\infty dr e^{-r} \right)= 4C_1^2 a^2 
\label{ctheorem3}
\ee
with $C_1, a$ appropriate constants~\cite{gang}.
Thus we have a quantitative description of the flow in Liouville time
towards flat space times :
$\lambda ^2 \rightarrow 0$ as $t \rightarrow \infty$. Could such
a mechanism also be operational in our four-dimensional world? If so,
how rapidly would the cosmological constant be relaxing, and how would its
magnitude compare with current experimental limits?

\subsection{$D$-Brane Foam and the Decoherent Scattering of Light
Particles}

As another example, we consider a string picture of space time-foam, i.e.,
microscopic quantum-gravity fluctuations, described as
$D$-branes, which provide
a very powerful representation of string solitonic states~\cite{dbrane}.
A conformal field theory formulation of soliton recoil in string theory,
has been developed in~\cite{km,kogwheat,lizzi,emnd},
exploiting the $D$-brane representation of the horizon
of black holes~\cite{dbrane} to describe the
back reaction of quantum fluctuations of matter on space time.

The basic observation of \cite{km,kogwheat} was that 
such recoil degrees of freedom can be described by a logarithmic conformal 
field theory~\cite{gurarie}. To see this, we recall that the
propagation of a light closed-string particle through this representation
of $D$-brane foam
involves, at the lowest order, a diagram with a disk topology, 
internal tachyon vertices, and 
Dirichlet boundary conditions~\cite{dbrane}. In this way one may describe 
scattering through a real or virtual $D$-brane state, with production
and decay amplitudes
$A_{TTm}$, subject to the energy-conservation condition. 
The next term in a topological expansion in genus
$g=2-2\#_{handles}-\#_{holes}$ is
an annulus with closed-string operator insertions. As has 
been discussed elsewhere~\cite{recoil,diffusion},
this has a singularity 
${\cal A} \sim  \delta (E_1+E_2)\sqrt{\frac{1}{\rm log(\delta )}}$
in the pinched-annulus configuration $\delta \rightarrow 0$, which is
regularized by introducing recoil 
operators~\cite{emnd,kogwheat,diffusion} $C,D$ 
to describe the back reaction 
of the struck $D$ brane:  
\be
 V_{rec} = y_i C + u_i D \qquad 
C \equiv \epsilon \int _{\partial \Sigma} \Theta_\epsilon (X^0) 
\partial_n X^i 
\qquad D  \equiv \int _{\partial \Sigma} X^0 \Theta_\epsilon (X^0) 
\partial_n X^i 
\label{CDpair}
\ee
where $y_i$ and $u_i$ are the 
position and momentum of the recoiling $D$ brane, and 
${\rm lim}_{\epsilon \rightarrow 0}\Theta _\epsilon (X^0)$ is a suitable 
integral representation of the step function~\cite{kogwheat}, with
$\epsilon$ a suitable infrared regulator parameter. 

The quantum treatment of $D$-brane recoil necessitates the
introduction of world-sheet annulus diagrams, whose
large-size limit is characterized by a
size parameter $L$ that, together with a conventional
ultraviolet world-sheet regulator parameter $a$, specifies
the value of $\epsilon$~\cite{kogwheat}:
\begin{equation}
{1 \over \epsilon^2} \simeq 2 \hbox{log} |L/a|^2
\label{epsilon}
\end{equation}
As discussed earlier, we identify the Liouville
field $\phi$ with the renormalization scale on the world 
sheet~\cite{emn,kogan},
and its zero mode, $\phi_0$, is further identified with the time variable
\begin{equation}
\phi_0 \, = \, t \, \simeq \, \hbox{log} | L / a|
\label{time}
\end{equation}
Note that there is no absolute time in this approach, since physical
quantities are described by the renormalization group,
which relates different scales $L, L'$ that correspond to time
differences $\delta t \simeq \hbox{log}|L/L'|$.
As discussed in~\cite{diffusion}, we
identify $1/\epsilon^2 \sim \hbox{log} \delta$, and in turn, using the
Fischler-Susskind mechanism~\cite{fs} 
on the world sheet to relate renormalization-group infinities 
among different genus surfaces, we identify $t \sim {\rm log}\delta$.   

The operators $C, D$ 
consitute a logarithmic pair~\cite{kogwheat} with 
$<C(z)C(0)>$, $<C(z)D(0)>$ 
non-singular as $\epsilon \rightarrow 0^+$, whereas   
$<D(z)D(0)>$ is singular with a world-sheet 
scale dependence~\cite{kogwheat} $D \rightarrow D + 
tC$, from which we infer that $u_i \rightarrow u_i$,
$y_i \rightarrow y_i + u_i t$, corresponding to a Galilean 
time transformation~\cite{lizzi}, as is appropriate 
for a heavy $D$ brane with mass $\propto 1/g_s$ (\ref{velocityoptimum}). 
This representation or recoil is a striking confirmation of the utility of
our identification of the Liouville field with time.

Continuing the analysis, one finds that
the logarithmic operators (\ref{CDpair}) 
make divergent contributions  to the genus-0 amplitude 
in the limit where it becomes a pair of Riemann surfaces 
$\Sigma_1, \Sigma_2$ connected by a degenerate strip~\cite{emnd,lizzi}:  
\bea
{\cal A}_{strip} & \sim & g_s ({\rm log}^2 \delta) \int d^2 z_1 D(z_1)
\int d^2 z_2 C(z_2), \nn \\ 
 &g_s& {\rm log}\delta\int d^2 z_1 D(z_1) \int d^2 z_2 D(z_2), \quad 
g_s{\rm log}\delta \int d^2 z_1 C(z_1) 
\int d^2 z_2 C(z_2)
\label{stripdiv}
\eea
Assuming a dilute gas of monopole defects on the world sheet, the amplitudes
(\ref{stripdiv}) become contributions to the effective 
action~\cite{emnd,lizzi}. 
One may then seek to cancel them or else to absorb them into scale-dependent 
couplings of the world-sheet field theory.
If these could all be cancelled, 
the corresponding theory would be conformally invariant, 
whereas absorption of these divergent contributions would result 
in departures from criticality. 

The leading double logarithm 
associated with the CD combination in (\ref{stripdiv}) 
may indeed be cancelled~\cite{lizzi} by imposing the momentum-conservation 
condition
\begin{equation}
u_i =  g_s(k_1 + k_2)_i  
\label{velocityoptimum}
\end{equation}
as expected for a $D$ brane soliton of mass $1/g_s$. 
As we now show, this
is also consistent with the tree-level energy-conservation 
condition that is obtained by integration 
over the Liouville zero mode, supporting its interpretation as time.
{}From the point of 
view of the Liouville theory on the open world sheet, 
the tree-level monopole 
mass term arises from a boundary term $i\int_{\partial \Sigma}
QX^0{\hat k}$ in the effective action~\cite{ambjorn}, where
${\hat k}$ is the extrinsic curvature 
and $Q$ is given by equation (\ref{31/2}). 
When the Liouville 
integration is performed at the quantum level, $Q$ is replaced
by its value in (\ref{31/2}), which receives
a contribution from $\beta_{y_i}=u_i$~\cite{diffusion}, by virtue of the 
logarithmic operator product expansion of $C$ and $D$~\cite{kogwheat}. 
Expanding the right-hand side of (\ref{31/2}) using (\ref{ctheorem}),   
for small $|u_iu^i| << 1$,  
we find the quantum energy-conservation condition:
\bea
E_1 + E_2 =\frac{e}{\sqrt{\beta}}=e\left(\pi(C-25)/3g_s\right)^{1/2}
= \frac{e}{\sqrt{g_s}}(1 + \frac{u_i^2}{2} + \dots )
\label{qecons}
\eea
which matches the momentum-conservation condition (\ref{velocityoptimum})
when we set $e\sqrt{\pi/3}=1/\sqrt{g_s}$. In this way, we
confirm our interpretation of time as the Liouville field~\cite{emn,kogan}. 
The fact that the cancellation of the leading double logarithm 
in (\ref{stripdiv}) enforces energy conservation $\frac{d}{dt}<E>=0$
confirms previous arguments~\cite{emn} in the general
renormalization-group approach to Liouville dynamics. 

The single logarithms associated with the CC and DD contributions 
in (\ref{stripdiv}) are a different story, since they 
can only be absorbed into quantum coupling parameters~\cite{emnd,lizzi}:
${\hat y_i}=y_i + \alpha_C\sqrt{{\rm log}\delta}, {\hat u}_i = u_i + 
\alpha_D\sqrt{{\rm log}\delta}$. 
The resulting probability distribution in theory space  
becomes time dependent~\cite{emnd,lizzi}:
\begin{equation}
{\cal P} \sim 
 \frac{1}{g_s^2log\delta} 
e^{-\frac{({\hat q}_m -q_m)G^{mn}({\hat q}_n - q_n)}{g_s^2log\delta}} 
\label{timprob}
\end{equation}
where $q_m \equiv \{ y_i,    u_i \}$. 

This in turn leads to a breakdown of the usual $S$-matrix
description of the transition from the 
initial-state density matrix. To see this, we first
use the Liouville dressing of the recoil operators~\cite{kogwheat},
and the identification of the Liouville mode with the time $X^0$ 
appearing already in the logarithmic pair (\ref{CDpair}),
to derive~\cite{emnd} the following
form for the singular part of the target-space metric $G_{MN}$
around the moment of the collision:
\begin{equation}
G_{00} = -1, \, G_{ij} = \delta_{ij}, \,G_{0i} = G_{i0} = \epsilon\,
(\epsilon y_i + u_i t) \Theta_{\epsilon} (t) 
\label{metric}
\end{equation}
It is to be understood that, in addition to (\ref{metric}), there
is also a static, spherically-symmetric part of the metric, which
is ${\cal O}(M / R)$ at large distances $R$ from the struck $D$ brane
of mass $M$. We now consider the scattering of a second low-energy
light particle at large impact parameter $R$, so that we may 
neglect this spherically-symmetric part
in a first approximation, and consider the asymptotic
metric as flat to zeroth order in $M / R$. The physics that interests
us is that associated with the $\epsilon$-dependent singularity in
(\ref{metric}).

\begin{centering}
\begin{figure}[htb]
\epsfxsize=2in
\centerline{\epsffile{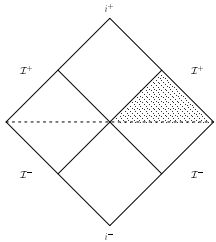}}
%
%
\caption{{\it Penrose diagram for the 
space-time environment derived from the
quantum recoil of a heavy $D$ brane induced by the
scattering of a light closed-string state.}}
\label{rindlerfig}
\end{figure}
\end{centering}

There is particle creation associated with the metric (\ref{metric}).
To see this, we first note that an on-mass-shell
scalar field $\phi $ of mass $\mu$ in the background (\ref{metric})
may be expanded in terms of ordinary flat-space Minkowski modes which
play the r\^ole of the ``in'' modes, since
our background space (\ref{metric}) can be
mapped, for $t > 0$, to a flat space to ${\cal O}(\epsilon ^2)$
by means of a simple coordinate transformation:
\be 
{\tilde t} = t \qquad, 
\qquad {\tilde X}^i = X^i + \frac{1}{2}\epsilon  u_i t^2
\label{rindler2}
\ee
which may be represented by the Penrose diagram shown
in Fig. \ref{rindlerfig}. 

Each of the four diamonds corresponds in a conventional
Rindler space to one of the causally-separated regions~\cite{birrel},
and our space-time corresponds to the right-hand diamond.
We see that it is flat
Minkowski for $t < 0$, and that for $t >> 0$ 
the shaded region of the space-time resembles, to ${\cal O}(\epsilon )$,
a Rindler wedge, with `acceleration'  $\epsilon u_i$. The event horizons
are indicated by the straight lines separating the diamond-shaped 
regions of the diagram. The dashed line corresponds to the 
curvature singularity $\epsilon ^2 \delta_{\epsilon} (t)$, which may be
ignored in order ${\cal O}(\epsilon)$.   
As a consequence of the non-relativistic nature of the heavy $D$ brane,
considered here,  
this singularity  appears to violate causality, lying outside the 
light cone. This is merely an artifact of taking the
limit as the velocity of light 
$c \rightarrow \infty$, as is appropriate for a non-relativistic (heavy) 
$D$ particle. One expects causality to be restored, with a rotation of the 
singular surface so as to become light-like, in a fully relativistic 
treatment of the problem, which lies beyond the scope of the present 
discussion.

Analyzing the non-trivial
Bogolubov transformation between the ``in" and ``out" vacua
at $t < 0$ and $t >> 0$, we find~\cite{ekmnw}
the following number of particles created in the mode labelled by $E$ and
$K$:
\be
n_{EK} \propto {\frac{\epsilon ^2 ({\underline u}.{\underline
K})^2}{E^6}}
\label{creation2}
\ee
Note that the formula (\ref{creation2}) 
describes particle creation with a non-trivial angular
distribution around the direction of the velocity vector
$u_i$. The particle spectrum is {\it not thermal} as it
would have been in the case of a uniform acceleration~\cite{unruh}.

It can be shown that the pertinent 
influence action $S_{IF}$ for the scale field~\cite{hu}
is, in order of magnitude, given by: 
\bea
S_{IF} \propto -i \hbar {\rm log}{\cal F}(a,a') \sim 
\sim i\hbar 
[\frac{1}{4}g_s^2 |{\underline {\hat u}}.{\underline K}|^2~t  + 
 \dots 
\label{final}
\eea
where ${\underline {\hat u}}$ 
denotes a unit vector in the direction of ${\underline u}$, and 
the $\dots $ indicate terms
which are not of interest to us here.
Thus the quantum fluctuations of the $D$-brane recoil velocity,
reperesented by the summation over world-sheet
genera~\cite{diffusion,lizzi}, 
induce decoherence for the second light particle that grows linearly in
time. If there were only the classical recoil velocity,
there would be no ${\rm log}\delta$ dependence
and hence no time dependence and no decoherence.

The imaginary coefficient in (\ref{final}), and its linear
dependence on $t$, imply that the quantum recoil of the $D$ brane,
which corresponds in conformal-field-theoretic language to a
cenario
departure from criticality, induces 
a non-Hamiltonian contribution $\nd{\delta H}$ (\ref{ehns})
into the generic
evolution equation of the reduced density matrix for
the scalar particle in this background. 
Since 
the momentum of the 
spectator particle, $K$, may be taken generically to be of the same order 
of magnitude as  
its energy $E$,  
one obtains from (\ref{final})
the order-of-magnitude estimate
\be 
\nd{\delta H} \sim {\cal O}( E^2/M_D)
\label{hslash}
\ee
where $M_D$ is the heavy $D$-brane mass, which is related to the 
conventional string scale $M_s=(\alpha ')^{-1/2}$, with $\alpha'$ the Regge 
slope, via $g_s \sim M_s/M_D < 1$. 
The formula (\ref{hslash}) agrees
with the generic estimates of~\cite{ehns,emn,emnd},  
with suppression by only a single power of the heavy mass scale, 
in this case $M_D$.

\section{What is the Future of Time?}

Clearly, the ideas discussed in this review are often heuristic,
though based on plausible intuition and some detailed analysis in
specific models. Much work is needed to put the results on a
firmer theoretical foundation. We are of the firm opinion that
string theory~\cite{string}, in some non-perturbative incarnation such as
$M$ theory, provides the most appropriate framework for this
necessary work, and the technology of $D$ branes~\cite{dbrane}
is certainly a very powerful tool to tackle it, as we have attempted
in~\cite{emnm}. We do not attempt here to
preview future theoretical developments. However, we do believe
that one should constantly keep in mind possible experimental
probes of such speculative theoretical ideas, and conclude this
article by reviewing some tentative suggestions in this direction.

\subsection{Uncertain Times} 

The canonical quantization of string theory space has
been discussed in~\cite{emninfl,emnd}. Motion in
this space is characterized as a renormalization-group flow in
the space of effective $\sigma$-model couplings $g^i$ on the
lowest-genus world sheet, with the flow variable identified
in turn with the Liouville field and the target time variable.
This flow is classical in the absence of higher-genus effects.
However, in their presence, the renormalization-group flow has been
shown to obey the relevant Helmholtz conditions which are
necessary and sufficient for the
canonical quantization of the $\sigma$-model couplings $g^i$. This
potentially has interesting consequences for cosmology and provides a
possible string scenario for inflation and entropy
generation~\cite{emninfl}.

This summation over genera and the ensuing quantization clearly entail a 
modification of the conventional conception of space-time coordinates, though 
the exact nature of the resulting picture is still to be understood. Some 
aspects are, however, already apparent. It is clear from the
identification of target time $t$ with the zero mode of the Liouville
field, up to a $g^i$-dependent normalization factor:
\be
t=Q[g^i] \, \varphi_0 \qquad Q[g^i]^2 \equiv \frac{1}{3}({\cal C}[g^i]-25)
\label{target}
\ee
where $\varphi_0$ is the world-sheet zero mode of the Liouville scale
$\varphi$ in (\ref{Lscale}), and ${\cal C}[g^i]$ the
Zamolodchikov ${\cal C}$-function, that $t$ also becomes a quantum
operator.
The identification of the target-space coordinates
as zero modes of the $\sigma$-model fields $X$ also
implies in non-critical strings an interdependence between these
coordinates and the quantum background $\{ g^i \}$. 
Although the precise 
nature of the (correspondingly) quantized target-space
coordinates remains to be fully understood,
these observations suggest the appearance of
non-trivial space-time commutation and uncertainty relations. 

This can be seen somewhat more directly
in the dynamical context of D branes treated as non-relativistic heavy
objects. 
The summation over genera in that case can be shown~\cite{amelino} 
to yield an uncertainty relation between the 
Liouville time and transverse spatial coordinates 
of the $D$-brane: 
\be
 \delta y \delta t \, \ge \, 
\sqrt{g_R} \alpha '
~, \label{dxdtaemn}
\ee
where $g_R$ is an appropriately renormalized~\cite{lizzi} 
string coupling, and $\alpha '$ is the Regge slope.  
Uncertainties resembling (\ref{dxdtaemn}) have also been
derived in the context of critical $D$ branes~\cite{yoneya}, but 
in that case the right-hand side depends only on the 
Regge slope $\alpha'$, and not on the string coupling.  

These features may lead to uncertainty $\delta t$ 
in the time delay of light
propagation, leading to an energy-dependent dispersion in the apparent
velocity of light, which might take the form ${\delta c \over c} \sim ({E
\over M})$, where $M$ is some gravitational mass scale. The most sensitive
probe of such a possibility may be provided by gamma-ray bursters (GRBs),
some of which are known to be at cosmological distances, and whose
emissions may have short-time structures. We have estimated~\cite{sarkar} 
that a
careful observational programme might be sensitive to $M \sim 10^{18}$
GeV, comparable to the possible mass scale of quantum gravity.

\subsection{Towards Experimental Tests of Decoherence
Induced by Quantum Gravity}

Tests of quantum-gravity entanglement may be possible in 
particle physics, using sensitive probes of quantum mechanics.
The most sensitive probe, to date, is the neutral kaon system. 
Assuming~\cite{ehns} energy and probability 
conservation, it was shown that
one can parametrize the modification term $\nd{\delta H}$ 
in (\ref{ehns}) as~\cite{ehns}
\begin{equation}
  {\nd h}_{\alpha\beta} =\left( \begin{array}{c}
 0  \qquad  0  \qquad   0      \qquad     0  \\
 0  \qquad  0  \qquad   0      \qquad     0  \\
 0  \qquad  0  \qquad -2\alpha \qquad   -2\beta  \\
 0  \qquad  0  \qquad -2\beta  \qquad   -2\gamma 
\end{array}\right)
\label{deltah}
\end{equation}
where the indices $\alpha, \beta$ label Pauli matrices
$\sigma_{\alpha, \beta}$ in the $K_{1,2}$ basis, and we
have also assumed that $\nd{\delta H}$ has $\Delta S = 0$. 
The three free parameters $\alpha,\beta,\gamma$
must obey the conditions
\begin{equation}
\alpha , \, \gamma \, > \, 0, \qquad \alpha \gamma \, > \, \beta^2
\label{positiv}
\end{equation}
stemming from the positivity of the matrix $\rho$. 

It is easy to see that these parameters 
induce decoherence and violate CPT~\cite{ELMN}. 
Various observables sensitive to these parameters have
been discussed in the literature, including 
the $2 \pi$ decay asymmetry
and the double semileptonic decay asymmetry.
Measurements of these and other quantities would 
in principle enable the decohering CPT violation
that we propose here to be distinguished
from `conventional' quantum-mechanical CPT violation~\cite{ELMN}.
The data so far agree perfectly with conventional
quantum mechanics, imposing only the following upper limits~\cite{emncplear}:
\begin{equation}
\alpha < 4.0 \times 10^{-17} \hbox{GeV}, \qquad 
\beta < 2.3 \times 10^{-19} \hbox{GeV}, \qquad 
\gamma < 3.7 \times 10^{-21} \hbox{GeV}
\label{bounds}
\end{equation}
We cannot help being impressed that these bounds are in the
ballpark of $m_K^2 / M_P$, which is the maximum 
magnitude that we could expect any such effect to have.

\pr
This and the previous example give us hope that 
experiments may be able to probe close to the Planck scale,
and we would not exclude the possibility of being able to
test some of the speculative ideas about quantum gravity reviewed
in this article.

\section*{Acknowledgements} 

The work of D.V.N. is supported in part by D.O.E. grant DE-F-G03-95-ER-40917.

\end{document}